\documentclass[a4paper,11pt]{article}
\pdfoutput=1 %

\usepackage{jinstpub} %
\PassOptionsToPackage{hyphens}{url}
\appto\UrlBreaks{\do\/}
\appto\UrlBigBreaks{\do\/}

\usepackage{lineno}
\usepackage{cprotect}
\newcommand{\wzonline}[1]{}

\newcommand{\wzurl}[1]{\href{#1}{\url{#1}}}

\title{\boldmath QEMU-based hardware/software co-development for DAQ systems}

\author{Wojciech M. Zabołotny}

\affiliation{Institute of Electronic Systems, Warsaw University of Technology,\\
    Nowowiejska 15/19, 00-665 Warszawa, Poland}

\emailAdd{wzab@ise.pw.edu.pl}

\abstract{Modern DAQ systems typically use the FPGA-based PCIe cards to concentrate and deliver the data to a computer used as an entry node of the data processing network.
	This paper presents a QEMU-based methodology for the co-development of the FPGA-based hardware part, the Linux kernel driver, and the data receiving application. \textcolor{black}{This} approach enables quick verification of the FPGA firmware architecture, organization of control registers, the functionality of the driver, and the user-space application.
	The developed design may be tested in different emulated architectures with a changeable CPU, IOMMU, size of memory, and the number of DAQ cards.}


\arxivnumber{2109.14735} %

\proceeding{TWEPP 2021\\
 20 -- 24.09.2021}

\begin{document}
\maketitle
\flushbottom

\section{Introduction}
\label{sec:intro}
Modern DAQ systems typically use the FPGA-based PCIe cards to concentrate and deliver the data 
to a computer being an entry node of the data processing network.
\textcolor{black}{For example, the BM@N experiment~\cite{dementev_fast_2021}
plans to use the standard commercial PCIe board equipped with Virtex 7 FPGA in its data acquisition chain.
The high-performance programmable chip provides a high bandwidth channel for data transfer and a long-term possibility to update the data concentration and preprocessing algorithms -- an essential functionality in readout chains for High-Energy physics experiments.}

\section{Standard development problems}
\begin{figure}[htbp]
	\centering %
	\includegraphics[width=.8\textwidth]{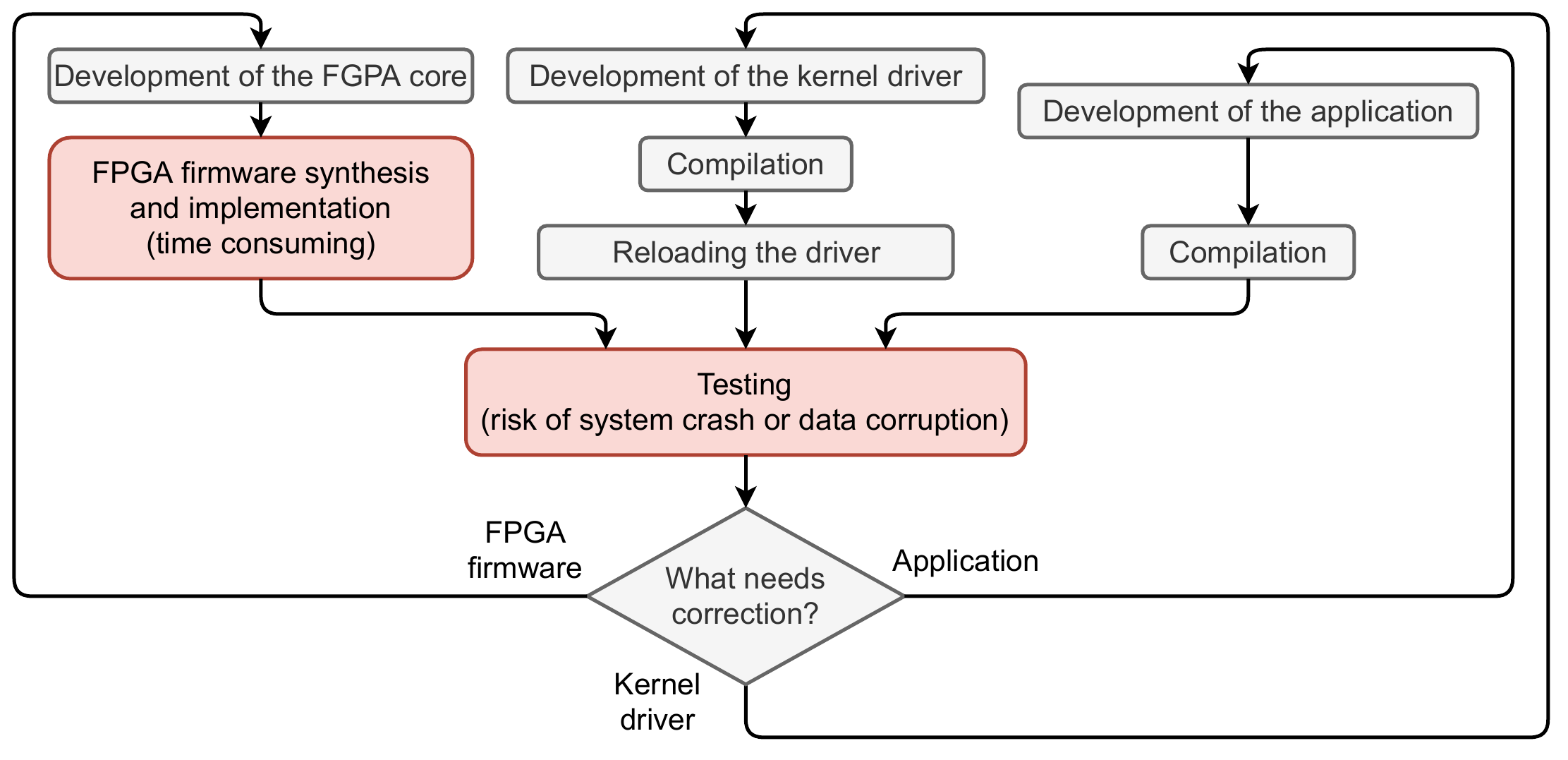}
	\caption{\label{fig:std-devel} The standard development cycle of the FPGA-based system with kernel driver and application.}
\end{figure}

The joint development of the FPGA firmware, the associated Linux kernel driver,
and the user-space application responsible for data reception and processing
is usually an iterative process with a long modification and testing cycle~\cite{bluj_physical_2016}.
\textcolor{black}{After modifying the FPGA code, the design must be synthesized and implemented, which may take
several hours for large FPGAs.
The DMA engines are difficult to debug.}
A bug in the bus-mastering DMA engine or its kernel driver may result in a system crash requiring a~reboot or even power-cycling
of the computer.
Sometimes, the crash may even lead to filesystem corruption requiring lengthy recovery \textcolor{black}{(see figure~\ref{fig:std-devel})}.

Another class of problems may be exposed when a few boards are used in the same computer and are controlled
by the same driver. \textcolor{black}{Failure to correctly separate the state of different devices or improper management of
system resources is a typical bug.}
Furthermore, testing the driver with multiple boards may be limited by hardware availability\textcolor{black}{, particularly early in the design cycle.
The possibility of testing the device and the driver on various hardware platforms may also be affected by that factor.}

Therefore, even though testing the design in the actual FPGA on the target hardware platform remains
the essential final testing step, a faster and less hardware-hungry method is desired during the development.

\section{Proposed method}
\label{sec:method}
The impact of possible system crashes may be minimized by using a virtual or emulated machine.
The filesystem image may be stored in a file on the host's disk, and \textcolor{black}{a copy} may be preserved for a quick recovery.
Many computer emulators, like QEMU~\cite{url-qemu} or 
OVPsim~\cite{url-ovpsim}\textcolor{black}{\footnote{
The significant disadvantage of OVP is that it is not fully open source. There are open source models of certain processors and peripherals, but the OVPsim simulator itself is closed and free only for non-commercial usage.}},
may emulate various hardware platforms and enable
easy modification of the parameters of the emulated machine.

A hardware simulator may be used to simulate the FPGA part.
However, the integration of the hardware simulator with an emulated machine is not trivial.
Solutions enabling HDL co-simulation are rare~\cite{url-texane,cho_full-system_2018}.
Most available solutions~\cite{4374971,5678362,cucchetto_common_2014} use SystemC. %
However, it is not a widely used language for the description of FPGA-based systems.
The Xilinx VivadoHLS tool supported SystemC-based synthesis, 
but support was dropped in version 2020.2~\cite{url-xlx-drop-systemc}.
Usually, SystemC requires relatively expensive
third-party tools like Catapult C~\cite{url-catapultc} or Stratus HLS~\cite{url-stratus-hls}.

Even though most co-simulation solutions correctly simulate the response of the FPGA part
to requests coming from the emulated machine (like read or write operation), 
they cannot correctly simulate the emulated machine's response to an event generated \textcolor{black}{by the FPGA}.
The latter is necessary to simulate interrupts, \textcolor{black}{and a DMA write} to the machine memory.
\textcolor{black}{For these purposes}, tight synchronization between both simulators is required~\cite{zabolotny_development_2012},
significantly slowing down the computer emulation. 
An interesting solution to that problem is proposed in~\cite{cho_full-system_2018}, but unfortunately,
the authors do not publish sources.

\textcolor{black}{With the above facts, an alternative approach is to create a model of the FPGA-based device}
in the open-source QEMU emulator in C language.
Of course such implementation cannot be directly used for logic synthesis, but on the other hand,
it may be easier to understand and modify by software engineers developing the driver and application.

\textcolor{black}{QEMU provides a set of functions and an example~\cite{url-qemu-edu-dev} to create hardware device models.}

\section{Practical application - a DMA engine}

\begin{figure}[htbp]
	\centering %
	\includegraphics[width=.7\textwidth]{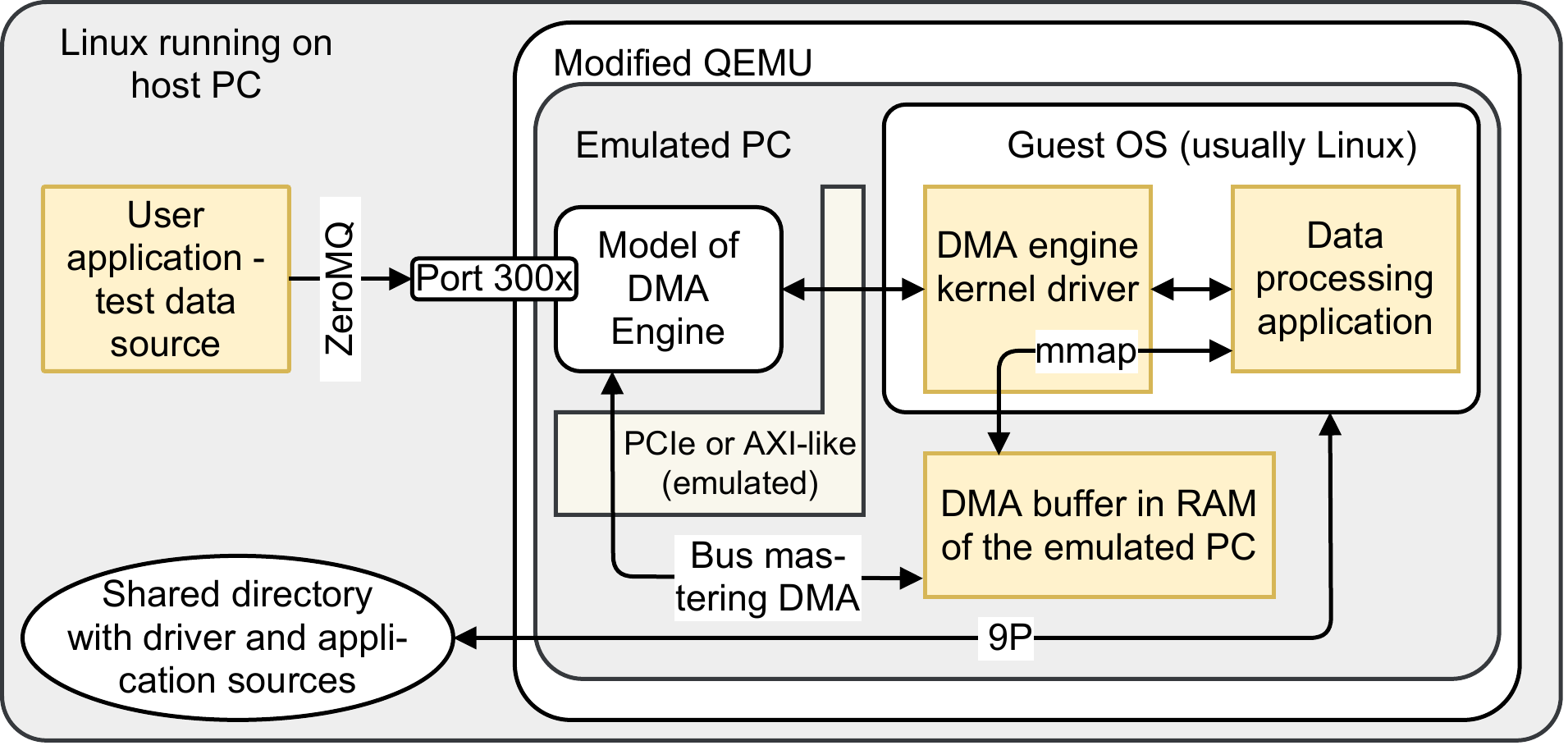}
	\caption{\label{fig:bd-sol} The system used for the development and testing
		of the virtual prototype of the DMA engine. \textcolor{black}{The model of the DMA engine is a part of the QEMU code. It receives test data via the ZeroMQ protocol from an application running on the host PC. It is also connected to the emulated system bus (PCIe or AXI-like). The engine is a bus slave controlled by the driver and works as a bus master writing the data to the DMA buffer in the RAM of the emulated machine. The DMA buffer is mapped into memory of the data processing
application using the mmap function of the driver.}}
\end{figure}

\textcolor{black}{The method below} has been used to develop and test the virtual prototype of the DMA-enabled data concentrator. 
\textcolor{black}{The block diagram of the proposed solution is shown in figure~\ref{fig:bd-sol}.}
The aim was to create a PCIe or AXI-connected bus mastering device that receives the concentrated data via an AXI-Stream interface
and stores it in the circular buffer, in which it is directly accessible for the data processing application.

The device should, as much as possible, avoid legacy dependencies.
 For example, the intention was to use the big DMA buffers allocated
\textcolor{black}{at run-time} by the user-space application and avoid solutions
not available in distribution kernels, like CMA. 
Therefore the scatter-gather approach was necessary. Unfortunately, the 1~GB DMA buffer consists of \textcolor{black}{262,144 4~KB} pages 
on most platforms. Storing the physical address of each page in the FPGA is a waste of resources. Storing them
in the computer memory and reading them \textcolor{black}{by DMA} is a waste of bandwidth.
 Therefore, the hugepage-backed
buffers were used. For a 1~GB DMA buffer, only 512 2~MB pages are needed, so their physical addresses may be 
easily stored in a BRAM inside FPGA.

Usage of scattered hugepage-backed application-allocated buffers for DMA
requires usage of Linux kernel APIs that are still being developed. 
The communication between the FPGA part, the kernel driver, and the application requires
thorough optimization regarding the performance, ease of use, and reliability.
The DMA engine is also difficult to debug in the real hardware, as bugs in the FPGA part
or in the kernel driver easily result in serious system crashes.

The listed features made the project an ideal candidate for using the method described in section~\ref{sec:method}.

The model of the DMA engine has been implemented in C as a part of modified QEMU~\cite{url-wzab-wzdaq}.
Of course, testing of the engine requires the test data.
 The data may be generated by an application running on the QEMU host
  and delivered to QEMU via the ZeroMQ-based protocol to avoid unnecessary QEMU-related overhead.
   In the case of \textcolor{black}{multiple instantiated devices}, they receive data at consecutive ports, starting from port 3000.

The kernel driver and test applications are available in the GitLab repository~\cite{url-wzdaq-driver}.
The kernel driver may handle multiple DMA engines (the maximum number is defined \textcolor{black}{before compilation}).

\section{Results}
\textcolor{black}{The DMA engine model} has been added to the {\bf i386} and {\bf ARM virt} machines and successfully compiled for {\bf x86\_64\_softmmu} and {\bf aarch64\_softmmu} targets.
The resulting QEMU versions were successfully run with the Debian Linux system installed on the virtual machines.
The disk images created during installation were copied to the backup files for quick recovery.

For testing, the required number of emulated concentrator devices was added using the appropriate number of 
\verb|-device pci-wzdaq1,dsn=N| options (where N is the unique ``device serial number'') in the invocation of QEMU.

\textcolor{black}{The sources for the application} and driver were stored and edited on the host machine,
shared via 9p protocol with the emulated machine, and compiled natively on the emulated machine.

The boot time of the emulated system was below 20 seconds (on a host with Intel i7-4790 CPU), enabling very quick crash recovery. 
Recompilation of each component - the QEMU model of the device, the driver, and the application consumed less than 1 minute, which enabled a quick development cycle.
During the development, a few catastrophic errors in the device model and in the kernel driver were identified and fixed.
 Many of them resulted in a crash of the emulated system.
The use of the proposed method resulted in a significant speedup of the development
 via a faster reboot of the test machine and a much faster modification-testing cycle of the DMA engine.

The possibility to move the created system to another Linux machine was also tested and confirmed. 
The QEMU enabled easy modification of the virtual machine parameters (changing the number of CPU cores, amount of RAM, and other parameters).

The created DAQ system was successfully tested on an emulated machine with 16 GB of RAM (the host had 32 GB of RAM) with eight simulated DAQ boards. Each board used a 1 GB data buffer consisting of 512 hugepages with a size 2 MB each.
The data were delivered with ZeroMQ via a local TCP/IP socket from the local data-generating applications.
\textcolor{black}{The obtained data throughput was approximately 3.22~GB/s for a single emulated board and 2.38~GB/s for eight simulated boards\footnote{\textcolor{black}{Of course, the throughput is highly dependent on the performance of the host computer and the implementation of the data source and data processing application.}}.} 

\section{Conclusion and future plans}
The proposed method enables quick development and early verification of the FPGA-destined
hardware concept together with the accompanying kernel driver.
With that approach, multiple developers may simultaneously develop the device model,
driver, and application, testing them on their computers.
The development may be started before the hardware is available and may even help
select the FPGA platform.
The results suggest that the proposed methodology may be a valuable tool in developing
the new FPGA-based DAQ firmware.
The C-implemented model must be translated into the HDL description suitable for synthesis.
Further research is needed to investigate if that process may be automatized using the HLS technology.

\acknowledgments
The work has been partially supported by the statutory funds of Institute of Electronic Systems. This project has also received funding from the European Union’s Horizon 2020 research and innovation programme under grant agreement No 871072.

\bibliography{qemu_daq}
\bibliographystyle{JHEP}

\end{document}